\documentclass[nofootinbib,aps,prb,reprint,superscriptaddress,twocolumn, draft=false]{revtex4-1}
\usepackage{hyperref}
\hypersetup{colorlinks=true,
		    linkcolor=blue,
		    citecolor=blue,
		    allcolors=blue}
\usepackage{amsmath, amsfonts, bm, bbm, braket, mathtools}
\usepackage{graphicx}
\usepackage{natbib}
\usepackage{lipsum}	
\usepackage{comment}	    

\newcommand{\norm}[1]{\left\lVert#1\right\rVert} 
\newcommand{\pmat}[1]{\begin{pmatrix} #1 \end{pmatrix}} 
\renewcommand{\d}{\mathrm{d}}
\newcommand{\id}{\mathbbm{1}}
\renewcommand{\H}{\mathcal{H}}
\newcommand{\V}{\mathcal{V}}
\newcommand{\e}{\xi}
\newcommand{\eh}{\varepsilon}

\newcommand{\vp}{\varphi}
\newcommand{\w}{\omega}
\newcommand{\dv}{\bm{d}}
\renewcommand{\k}{\bm{k}}
\newcommand{\q}{\bm{q}}
\newcommand{\s}{\sigma}
\newcommand{\qa}{\q_A}
\newcommand{\qb}{\q_B}
\newcommand{\tk}{\theta_{\k}}
\newcommand{\pk}{\varphi_{\k}}
\newcommand{\nh}{\bm{\hat{n}}}
\newcommand{\zh}{\bm{\hat{z}}}
\newcommand{\ad}[1]{a^{\dag}_{#1}}
\renewcommand{\a}[1]{a^{\phantom{\dag}}_{#1}}
\newcommand{\hd}[1]{h^{\dag}_{#1}}
\newcommand{\h}[1]{h^{\phantom{\dag}}_{#1}}
\newcommand{\cd}[1]{c^{\dag}_{#1}}
\renewcommand{\c}[1]{c^{\phantom{\dag}}_{#1}}
\newcommand{\gd}[1]{\gamma^{\dag}_{#1}}
\newcommand{\g}[1]{\gamma^{\phantom{\dag}}_{#1}}

\begin{document}

\title{Anisotropic Purity of Entangled Photons From Cooper Pairs in Heterostructures}

\author{Jacob S. Gordon}
\affiliation{Department of Physics, University of Toronto, Ontario M5S 1A7, Canada}
\author{Hae-Young Kee}
\email{hykee@physics.utoronto.ca}
\affiliation{Department of Physics, University of Toronto, Ontario M5S 1A7, Canada}
\affiliation{Canadian Institute for Advanced Research, CIFAR Program in Quantum Materials, Toronto, ON M5G 1M1, Canada}
\date{\today}
\begin{abstract}
It was theoretically proposed that a forward-biased $p$-$n$ junction with a superconducting layer (P-N-S) would produce pure, polarization-entangled photons due to inherent spin-singlet pairing of electrons. However, any heterostructure interface generically induces Rashba spin-orbit coupling, which in turn generates a mixed singlet-triplet superconducting order parameter. Here we study the effect of triplet pairing on the purity of photons produced through Cooper pair recombination. A unique directional dependence of the state purity is found for a triplet superconductor with fixed $\dv$: pure, entangled photons are produced when the photon polarization axis is parallel to $\dv$. Induced triplet pairing in a singlet superconductor is shown to degrade the state purity, while induced singlet pairing in a triplet superconductor is shown to enhance the production of entangled pairs. These considerations may aid the design of functional devices to produce entangled photons. 
\end{abstract}
\maketitle

\section{Introduction\label{sec:intro}}
Entanglement is a uniquely quantum mechanical phenomenon, which has found application in the rapidly growing fields of quantum computing\cite{raussendorf2001one,raussendorf2003measurement,obrien2007optical,ladd2010quantum}, cryptography\cite{ekert1991quantum,bennett1993teleporting,bouwmeester1997experimental,gisin2002quantum}, and metrology\cite{mitchell2004super,nagata2007beating}. Photons are a promising medium for encoding entanglement\cite{jennewein2000quantum}, and may be suitable for transportation of qubits\cite{ursin2004communications,ursin2007entanglement} with application in quantum key distribution. These applications require efficient sources of entangled photon pairs, and recent research has focused on mechanisms for their generation.

Widespread techniques for generating entangled photon pairs include parametric down-conversion in non-linear crystals\cite{shih1988new,ou1988violation,kwiat1995new,kwiat1999ultrabright,white2001exploring}, 
cascaded emission from biexcitons in semiconductor quantum dots\cite{benson2000regulated,stevenson2006semiconductor,akopian2006entangled}, and resonant hyperparametric scattering in semiconductors through a non-linear optical process\cite{edamatsu2004generation,oohata2007photon,nakayama2007observation}. In addition, hybrid devices incorporating semiconductors and superconductors have been proposed, including quantum dots wherein injected Cooper pairs simultaneously recombine\cite{suemune2006superconductor,de2010hybrid,khoshnegar2011entangled,nigg2015detecting}.

\begin{figure}[h]
	\includegraphics[width=\linewidth]{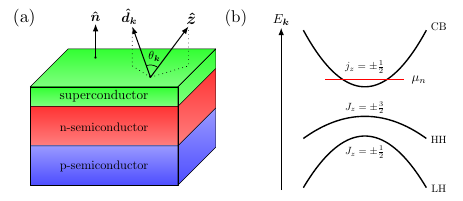}
	\caption{(colour online) (a) Schematic P-N-S heterostructure, consisting of a $p$-type semiconductor and an $n$-type semiconductor in which pairing is induced by the superconductor through the proximity effect. The heterostructure normal direction is defined by $\nh$; the photon polarization axis, by $\zh$; and a particular $\dv(\k)$ vector direction, by $\bm{\hat{d}}_{\k}$. The angle between $\zh$ and $\bm{\hat{d}}_{\k}$ is taken to be $\theta_{\k}$. (b) Schematic band structure of the semiconductors in the quantum well structure, showing the splitting of heavy-hole and light-hole bands.}
	\label{fig:heterostructure}
\end{figure}

Recently, there has been movement towards the use of two-dimensional semiconductor structures,  widely used in existing optoelectronic devices, for generating entangled photons.  An advance towards this technology has been proposed\cite{hanamura2002superradiance,hayashi2008superconductor} in the form of a P-N-S heterostructure, shown in Fig.~\ref{fig:heterostructure}(a), consisting of an ordinary $p$-$n$ junction with superconductivity induced in the $n$-type semiconductor through the proximity effect. Such a device exhibits enhanced electronically driven luminescence due to radiative Cooper pair recombination\cite{hayashi2008superconductor,asano2009luminescence,sasakura2011enhanced,hlobil2015luminescence}. In addition, it has been theoretically proposed \cite{hayat2014cooper} that with a quantum well structure of the semiconductor layers, it may also produce pure, polarization-entangled photon pairs. This phenomenon reflects the entanglement of electron spins in singlet Cooper pairs.
 
However, in such a heterostructure there is Rashba spin-orbit coupling\cite{bychkov1984properties,manchon2015new,winkler2003spin} (SOC) due to breaking of inversion symmetry across the interface. It is known that the presence of Rashba SOC generically leads to a mixed singlet-triplet superconducting order parameter\cite{edel1989characteristics,gorkov2001superconducting,yip2002two,edelstein2003triplet}, which has not been accounted for in previous studies\cite{asano2009luminescence,hayat2014cooper,hlobil2015luminescence}.

Here we study the effect of both singlet and triplet Cooper pairs on the purity of entangled photons produced in a P-N-S heterostructure with Rashba SOC. Triplet pairing breaks the rotational invariance of the Cooper pair spin state, which results in a unique directional dependence of the photon state purity. In a pure triplet superconductor with fixed $\dv$ vector, pure, entangled photons are produced when the photon momenta $\q$ are parallel to $\dv$, with weakened purity for other directions. We also show that the induced triplet pairing in a singlet superconductor hinders the production of pure, entangled photons due to directional dependence of the $\dv$ vector on $\k$. However, the induced singlet pairing in a triplet superconductor with fixed $\dv$ vector is shown to enhance the purity around $\q \parallel \dv$. These considerations may aid the design of compact and efficient devices for producing entangled photons.

This paper is organized as follows. We first review the P-N-S junction, interfacial Rashba SOC, and the induced component of the superconducting order parameter in Sec.~\ref{sec:induced-triplet}. We then calculate the two-photon density matrix for a pure spin-triplet superconductor and a singlet-triplet mixture in Sec.~\ref{sec:density-matrix}. From the density matrix, we obtain the photon state purity in Sec.~\ref{sec:purity} and demonstrate a unique angular dependence. Finally, discussion and material considerations are in the last section. Details of the calculation can be found in the Appendix. 

\section{Review of Induced Triplet Superconductivity in a P-N-S Junction\label{sec:induced-triplet}}
We consider a P-N-S heterostructure, shown in Fig.~\ref{fig:heterostructure}(a), which consists of a superconductor layer in electrical contact with the $n$-type semiconductor in an ordinary $p$-$n$ junction. A superconducting region is induced in the $n$-type semiconductor through the proximity effect\cite{degennes}. The semiconductors are taken to be of the zinc-blende type, such as GaAs or InP, which exhibit a direct band gap at the $\Gamma$ point. In bulk, the upper valence bands consist of heavy-hole (HH) and light-hole (LH) bands, which become degenerate at the zone center. The LH and HH bands have total spin-orbit-coupled momentum projections $J_z = \pm \tfrac{1}{2}$ and $J_z = \pm \tfrac{3}{2}$ ($\hbar \equiv 1$), while the conduction band (CB) has projection $j_z = \pm \tfrac{1}{2}$\cite{lee1974two}. In a quantum well structure, the degeneracy between the LH and HH bands is lifted\cite{bastard}, as shown in Fig.~\ref{fig:heterostructure}(b). Splitting of the LH-HH degeneracy has been shown\cite{hayat2014cooper} to be essential for the generation of pure, entangled photons, as Cooper pair recombination with the LH band degrades the state purity. Before calculating the two-photon density matrix, we first review the Rashba effect and the form of the induced pairing.

In a heterostructure with normal $\nh$, the surface-induced asymmetry gives rise to Rashba spin-orbit coupling of the form\cite{bychkov1984properties}\vspace{-0.5mm}
\begin{equation}
	\H_{\mathrm{Rashba}} = \lambda\sum_{\k}(\k\times\nh)\cdot\bm{\s}. 
\end{equation}
This leads to helicity-split bands with dispersion $\e_{\k,\pm} = \e_{\k} \pm \lambda \norm{\k}$, where $\k$ is understood to be a vector in the plane of the heterostructure. With inversion symmetry in the plane, the possible superconducting states can be classified based on parity. Cooper pairs are in a spin-singlet state for even orbital parity and a spin-triplet state for odd orbital parity \cite{mineev}. Rashba SOC breaks inversion symmetry within the plane, and generically, this will lead to a mixed singlet-triplet state of the Cooper pairs. 

Starting with a continuum model of the conduction band in the $n$-type semiconductor, $\xi_{\k} = \tfrac{k^2}{2m} - \mu_n$, where $\mu_n$ is the chemical potential, we focus on the limit that the energy splitting due to Rashba SOC is small compared to the Fermi energy
 \begin{equation}
 	\delta = \frac{m\lambda}{k_F}  = \frac{\lambda k_F}{2\mu_n} \ll 1.
 \end{equation}
This leads to a Fermi surface which is split as $k_{F,\pm} \approx k_F(1 \mp \delta)$. Assuming initial $s$-wave pairing of electrons near the Fermi surface, it has been shown with the method of Matsubara Green's functions\cite{gorkov2001superconducting} that both helicity-split bands acquire a superconducting gap at the Fermi surface,
\begin{equation}
	E_{\k,\pm} = \sqrt{\xi_{\k,\pm}^2 + |\Delta_0|^2},
\end{equation}
and that there is a spin-triplet admixture of Cooper pairs described by
\begin{equation}\label{eq:mixed-order-parameter}
	\hat{\Delta}_{\k}^{(m)} = \psi(\k)i\sigma_y + [\dv(\k)\cdot\bm{\s}]i\sigma_y,
\end{equation}
where $\psi(\k) = \Delta_0$ and $\dv(\k) \propto \bm{\hat{k}}$. The constant of proportionality is on the order of $\lambda$\cite{gorkov2001superconducting}.

If one instead starts with a pure triplet superconductor, then the Rashba spin-orbit coupling will induce a singlet component in the order parameter. The magnitude of the induced singlet component will also be on the order of $\lambda$.

Therefore, Cooper pairs in the P-N-S heterostructure exist as an admixture of spin-singlet and spin-triplet states through Rashba SOC. At each $\k$, the spin state of a triplet Cooper pair is described by the vector $\bm{d}(\k)$, with wave function
\begin{equation}
	\ket{\Psi_{\k}} \propto \bm{\hat{d}}(\k)\cdot[\cd{\k,\s}(\bm{\s}i\sigma_y)_{\s,\s'}\cd{-\k,\s'}]\ket{0},
\end{equation}
where $\cd{\k,\s}$ is an electron creation operator and repeated indices are summed over. It can be shown that\cite{coleman}
\begin{equation}\label{eq:spin-projection}
	\bm{\hat{d}}(\k)\cdot \bm{S}\ket{\Psi_{\k}} = 0,
\end{equation}
where $\bm{S}$ is the total spin operator. This means the $\dv$ vector is the direction on which the total spin of the pair has zero projection.

\section{Two-Photon Density Matrix \label{sec:density-matrix}}
In this section we calculate the two-photon density matrix using second-order time-dependent perturbation theory for a pure spin-triplet superconductor and a singlet-triplet admixture appropriate for the heterostructure. To model the $p$-$n$ junction, we take the unperturbed Hamiltonian
\begin{equation}\label{eq:unperturbed-H}
	\H_0 = \sum_{\q,\s}\w_{\q}\ad{\q,\s}\a{\q,\s} + \sum_{\k,J}\eh_{\k}\hd{\k,J}\h{\k,J} + \sum_{\k,j}\e_{\k}\cd{\k,j}\c{\k,j},
\end{equation} 
where $\ad{\q,\s},\hd{\k,J},\cd{\k,j}$ are creation operators for photons, holes, and electrons, respectively. The photon angular momentum runs over $\sigma = \pm 1$ ($\hbar \equiv 1$), corresponding to right/left ($R/L$) circular polarizations with respect to the quantization axis $\zh$, and we have neglected the zero-point energy. We will consider only the contribution from the HH band. As such, the $\zh$ projection of hole and electron angular momenta runs over $J_z = \pm \tfrac{3}{2}$ and $j_z = \pm \tfrac{1}{2}$, respectively. Once electron pairing is induced through the proximity effect, we adopt the BCS mean-field theory to describe the superconducting state
\begin{equation}\label{eq:bcs-H}
	\H_{ns} = \sum_{\k,j}E_{\k}\gd{\k,j}\g{\k,j},
\end{equation}
where $\gd{\k,j}$ is the creation operator for a Bogoliubov quasiparticle with energy $E_{\k} = \sqrt{\e_{\k}^2 + \Delta_{\k}^2}$.
The light-matter interaction responsible for the recombination process is 
\begin{equation}\label{eq:interaction-H}
	\V = \sum_{\k,\q,J,\s} B_{\k,\q}\ad{\q,\s}\h{\q-\k,-J}\c{\k,J+\s} + \mathrm{h.c.},
\end{equation} 
within the dipole approximation, where the parametrization has been chosen to conserve angular momentum along the photon axis $\zh$. Since we are considering the HH band only, the allowed angular momenta in the above interaction are $(\s,-J,J+\s) = (\pm 1, \pm \tfrac{3}{2}, \mp\tfrac{1}{2})$.

The initial state of the system is taken to be 
\begin{equation}\label{eq:initial-state}
\ket{\Psi_0} = \ket{0}\otimes\ket{\mathrm{FS}}\otimes\ket{\mathrm{BCS}},
\end{equation}
where $\ket{0}$ is the photon vacuum, $\ket{\mathrm{FS}}$ is the Fermi sea of holes in the HH band, and $\ket{\mathrm{BCS}}$ is the BCS state of electrons in the CB. In the interaction picture, the second-order contribution to the time-evolved state of the system is ($t_0 \rightarrow -\infty$)
\begin{equation}\label{eq:time-evolved-state}
\ket{\Psi_t} = \int_{-\infty}^t\d t_1\int_{-\infty}^{t_1}\d t_2\ \V(t_1)\V(t_2)\ket{\Psi_0},
\end{equation}
where $\V(t) = e^{i\H_0 t}\V e^{-i\H_0 t}$. To describe the state of the emitted photon pairs, we calculate the color-specific two-photon density matrix, with entries 
\begin{align}\label{eq:density-matrix-definition}
\begin{split}
	\rho_{\alpha\beta\gamma\delta}(\qa,\qb) &\coloneqq \braket{\Psi_t|\ad{\qa,\alpha}\ad{\qb,\beta}\a{\qa,\gamma}\a{\qb,\delta}|\Psi_t},
\end{split}
\end{align}
where $\alpha,\beta,\gamma,\delta\in\{\pm 1\}$. As discussed by \citet{hayat2014cooper}, there are one-photon emission processes in addition to the second-order two-photon emission processes from Cooper pair recombination. While the one-photon process produces photons with energy on the order of the band gap, $\w_{\q} = E_{\mathrm{BG}}$, the only constraint on photons emitted through the second-order process is that $\w_{\qa} + \w_{\qb} = 2E_{\mathrm{BG}}$. Therefore, the single-photon emissions can be distinguished from the two-photon emissions through spectral filtering; photons with energy differing from $E_{\mathrm{BG}}$ by more than the thermal broadening $k T$ must originate from Cooper pair recombination. This is the purpose of tracking the color dependence in the two-photon density matrix.

To understand the effect of Cooper pairs, the generalized Bogoliubov transformation for unitary pairing is employed\cite{sigrist1991unconventional}:
\begin{equation}\label{eq:bogoliubov-transformation}
\hspace{-2mm}\c{\k,j}(t) = e^{-i\mu_n t}\sum_{j'}\left[\hat{u}_{\k jj'} e^{-iE_{\k}t}\g{\k,j'} + \hat{v}_{\k jj'}e^{+iE_{\k}t}\gd{-\k,j'}\right],
\end{equation}
where
\begin{align}\label{eq:bogoliubov-matrices}
\begin{split}
	\hat{u}_{\k} &= \sqrt{\frac{1}{2}\left(1 + \frac{\xi_{\k}}{E_{\k}}\right)}\id_{\s}, \\
	\hat{v}_{\k} &= -\frac{\hat{\Delta}_{\k}}{\sqrt{2E_{\k}(E_{\k} + \e_{\k})}}.
\end{split}
\end{align}
Using the time-evolved operators, it is straightforward to calculate the expectation values appearing in Eq. (\ref{eq:density-matrix-definition}) using Wick's theorem. The contribution from a Cooper pair recombination process enters through terms of the form $\braket{\mathrm{BCS}|c^{\dag}c^{\dag}|\mathrm{BCS}}\braket{\mathrm{BCS}|cc|\mathrm{BCS}}$.

Once the matrix elements have been computed, the time integrals appearing in the time-evolved state (\ref{eq:time-evolved-state}) must be performed, which is done by adiabatically turning on the interaction $\V(t)$ from $t_0 \rightarrow -\infty$. This gives us the emission probability as a function of $t$, which we use to compute the emission rate\cite{landau2013quantum}. Details of this integration can be found in the Appendix. To distinguish the emission rate from probability, we will write $\overline{\rho}$ instead of $\rho$.  

To proceed, we focus on three cases of interest. First, we review the density matrix in the case of pure singlet pairing. Next, we consider intrinsic spin-triplet pairing and no singlet component in the order parameter. Finally, we consider the relevant case of a singlet-triplet mixture enabled through the heterostructure. 

\subsection{Pure Singlet Pairing}
A pure singlet superconductor has order parameter $\hat{\Delta}_{\k}^{(s)} = \psi(\k)i\s_y$, with gap $\Delta_{\k}^{(s)} = |\psi(\k)|$. In this case the matrix $\hat{v}_{\k}$ appearing in the Bogoliubov transformation Eq. (\ref{eq:bogoliubov-matrices}) takes the simple form
\begin{equation}\label{eq:bogoliubov-singlet}
	\hat{v}_{\k}^{(s)} = -v_{\k}^{(s)}\pmat{0 & -1 \\ +1 & 0},
\end{equation}
where $v_{\k}^{(s)} = \sqrt{\tfrac{1}{2}(1 - \xi_{\k}/E_{\k}^{(s)})}$. After calculating the electron expectations and performing the time integrals, we arrive at the density matrix
\begin{equation}\label{eq:singlet-density-matrix}
	\overline{\rho^{(s)}}(\qa,\qb) = \sum_{\k}F(\k,\qa,\qb,\Delta_{\k}^{(s)})\frac{1}{2}\pmat{0 & 0 & 0 & 0  \\ 0 & 1 & 1 & 0 \\ 0 & 1 & 1 & 0 \\ 0 & 0 & 0 & 0}, 
\end{equation}
where $F(\k,\qa,\qb,\Delta_{\k})$ is defined through Eq.~(\ref{eq:production-rate}) in the Appendix. The density matrix is written in the polarization basis $\{\ket{RR},\ket{RL},\ket{LR},\ket{LL}\}$ and corresponds to a pure, polarization-entangled state $\tfrac{1}{\sqrt{2}}(\ket{RL} + \ket{LR})$. \\

\subsection{Intrinsic Triplet Pairing}
In the case of an intrinsic spin-triplet superconductor, the order parameter is $\hat{\Delta}_{\k}^{(t)} = (\dv(\k)\cdot\bm{\s})i\s_y$ with gap $\Delta_{\k}^{(t)} = \norm{\dv(\k)}$. The direction of $\dv(\k)$ with respect to the photon polarization axis $\zh$ can be specified through the angles $(\tk,\pk)$
\begin{equation}\label{eq:dvec-angles}
	\frac{\dv(\k)}{\norm{\dv(\k)}} = (\sin\tk\cos\pk,\sin\tk\sin\pk,\cos\tk).
\end{equation}
It is important to note that due to the odd orbital pairing, $\dv(-\k) = -\dv(\k)$, the angles invert according to $(\tk,\pk) \mapsto (\pi-\tk,\pi+\pk)$, so that both $\cos\tk$ and $\sin\tk e^{\pm i\pk}$ change sign under inversion. The matrix $\hat{v}_{\k}$  can be rewritten as
\begin{equation}\label{eq:bogoliubov-triplet}
	\hspace{-2mm}\hat{v}_{\k}^{(t)} = -v_{\k}^{(t)}\frac{\hat{\Delta}_{\k}}{\norm{\dv(\k)}} = v_{\k}^{(t)}\pmat{+\sin\tk e^{-i\pk} & -\cos\tk \\ -\cos\tk & -\sin\tk e^{+i\pk}}, 
\end{equation}
where $v_{\k}^{(t)} = \sqrt{\tfrac{1}{2}(1 - \xi_{\k}/E_{\k}^{(t)})}$. This yields the density matrix
\begin{widetext}
\begin{equation}\label{eq:triplet-density-matrix}
\overline{\rho^{(t)}}(\qa,\qb) = \sum_{\k}F(\k,\qa,\qb,\Delta_{\k}^{(t)})\frac{1}{2}\begin{pmatrix}\sin^2\tk & 0 & 0 & 0 \\ 0 & \cos^2\tk & \cos^2\tk & 0 \\ 0 & \cos^2\tk & \cos^2\tk & 0 \\ 0 & 0 & 0 & \sin^2\tk\end{pmatrix},
\end{equation}
\end{widetext}
with the same definitions as in the singlet case and in the same polarization basis. The most important difference between the triplet and singlet cases are non-zero $\ket{RR}\bra{RR}$ and $\ket{LL}\bra{LL}$ components proportional to $\sin^2\tk$, vanishing in the limit that $\dv(\k)$ points along the photon polarization axis $\zh$. This effect is due to the existence of Cooper pairs with non-zero total spin projection on the $\zh$ axis, described by $d_{x} \pm i d_{y}$ in the order parameter. As we will see, these components reduce the purity of the photon polarization state, which acquires some directional dependence. \vspace{-0.5mm}

\subsection{Mixed Singlet-Triplet Pairing}
In the P-N-S heterostructure, we generally have a mixed singlet-triplet pairing of the form~(\ref{eq:mixed-order-parameter}),
\begin{equation*}
\hat{\Delta}_{\k}^{(m)} = \hat{\Delta}_{\k}^{(s)} + \hat{\Delta}_{\k}^{(t)} = \psi(\k)i\s_y + [\dv(\k)\cdot\bm{\s}]i\s_y,
\end{equation*}
with gap\cite{sigrist1991unconventional}
\begin{equation}
	\Delta_{\k}^{(m)} = \sqrt{\tfrac{1}{2}\mathrm{Tr}[\hat{\Delta}_{\k}^{(m)}\hat{\Delta}_{\k}^{(m)\dag}]} = \sqrt{|\psi(\k)|^2 + \norm{\dv(\k)}^2}.
\end{equation}
To account for the mixed pairing, it is useful to define quantities that measure the ``fraction'' of paring within the singlet and triplet channels, \vspace{-1mm}
\begin{equation}\label{eq:pairing-fraction}
	s_{\k} = \frac{|\psi(\k)|}{\sqrt{|\psi(\k)|^2 + \norm{\dv(\k)}^2}}, \quad 
	t_{\k} = \frac{\norm{\dv(\k)}}{\sqrt{|\psi(\k)|^2 + \norm{\dv(\k)}^2}},
\end{equation}
which satisfy $s_{\k}^2 + t_{\k}^2 = 1$. Using these quantities, we can decompose $\hat{v}_{\k}$ into singlet and triplet components,
\begin{equation}\label{eq:bogoliubov-mixed}
\hspace{-2.5mm}\hat{v}_{\k}^{(m)} = -v_{\k}^{(m)}\left[s_{\k}\pmat{0 & -1 \\ +1 & 0} + t_{\k}\pmat{-\sin\tk e^{-i\pk} & \cos\tk \\ \cos\tk & +\sin\tk e^{+i\pk}}\right]
\end{equation}
where $v_{\k}^{(m)} = \sqrt{\tfrac{1}{2}(1 - \xi_{\k}/E_{\k}^{(m)})}$. This leads to the density matrix 
\begin{widetext}
\begin{equation}\label{eq:mixed-density-matrix}
\hspace{-14mm}\overline{\rho^{(m)}}(\qa,\qb) = \sum_{\k}F(\k,\qa,\qb,\Delta_{\k}^{(m)})\frac{1}{2}\begin{pmatrix}t_{\k}^2\sin^2\tk & 0 & 0 & 0 \\ 0 & s_{\k}^2 + t_{\k}^2\cos^2\tk & s_{\k}^2 + t_{\k}^2\cos^2\tk & 0 \\ 0 & s_{\k}^2 + t_{\k}^2\cos^2\tk & s_{\k}^2 + t_{\k}^2\cos^2\tk & 0 \\ 0 & 0 & 0 & t_{\k}^2\sin^2\tk\end{pmatrix},
\end{equation}
\end{widetext}
which is a $\k$-dependent weighted sum of the pure singlet and triplet density matrices. In the case of induced triplet pairing in an $s$-wave superconductor, the superconducting gap on each helicity-split Fermi surface is actually $|\Delta_0|$\cite{gorkov2001superconducting} , so the function $F$ in Eq.~(\ref{eq:mixed-density-matrix}) must be evaluated at $\Delta_0$, with an extra factor
\begin{equation}
	\left|\frac{\Delta_{\k}^{(m)}}{\Delta_0}\right|^2.
\end{equation}
The extra factor is simply due to the fact that the gap is $|\Delta_0|$ rather than $\Delta_{\k}^{(m)}$, as assumed above. This form of the density matrix also holds in the case of induced singlet pairing in a triplet superconductor, with unspecified $\psi(\k)$.

\section{State Purity\label{sec:purity}}

\begin{figure}
	\includegraphics[width=\linewidth]{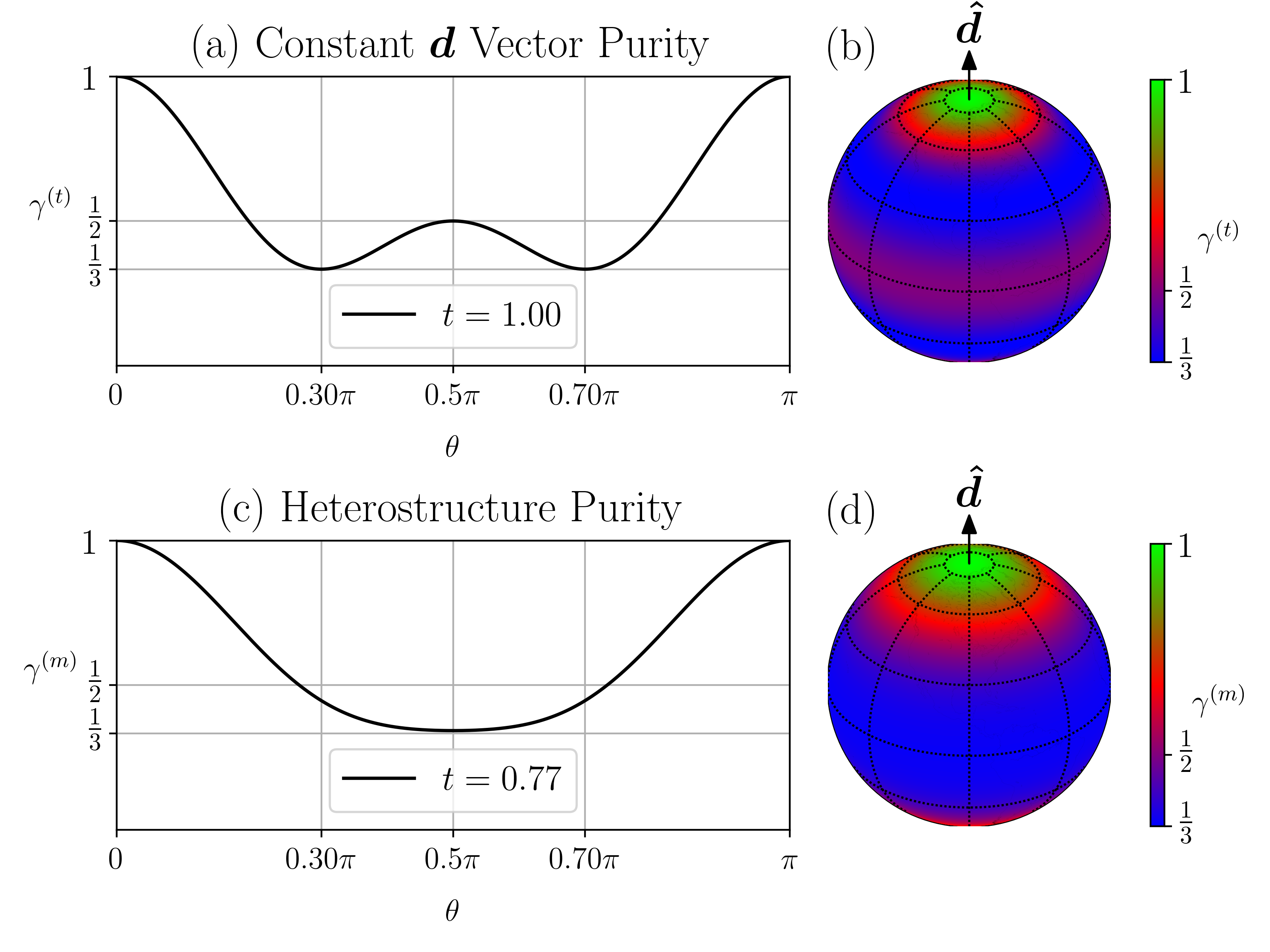}
	\caption{(colour online) (a) Calculated purity of photon pairs produced by spin-triplet Cooper pairs as a function of the polar angle $\theta$ between the photon polarization axis $\zh$ and the $\dv$ vector, and (b) full angular dependence. (c) Purity of photon pairs for triplet fraction $t = 0.77$ as a function of the polar angle $\theta$ between the photon axis $\zh$ and the $\dv$ vector, and (d) full angular dependence.}
	\label{fig:purity}
\end{figure}

We now turn our attention to the purity of the two-photon polarization state derived in the previous section. The function $F(\k,\qa,\qb,\Delta_{\k})$ outside the matrix describes the rate of production of photon pairs based on the occupation numbers, as well as the energies of the initial, intermediate, and final states in the second-order process. This is to be distinguished from the matrix itself, which describes the polarization state of the photons. As mentioned earlier, the polarization state of the photons produced by spin-singlet Cooper pairs is described by
\begin{equation}\label{eq:singlet-polarization}
	\rho^{(s)} = \frac{1}{2}\pmat{0 & 0 & 0 & 0 \\ 0 & 1 & 1 & 0 \\ 0 & 1 & 1 & 0 \\ 0 & 0 & 0 & 0},
\end{equation}
normalized such that $\mathrm{Tr}[\rho^{(s)}] = 1$. The state purity is easily seen to be $\gamma^{(s)} = \mathrm{Tr}[(\rho^{(s)})^2] = 1$ and is derived from the polarization-entangled state 
\begin{equation}\label{eq:photon-state}
	\ket{\Psi_{\mathrm{ph}}} = \frac{1}{\sqrt{2}}(\ket{RL} + \ket{LR}). 
\end{equation}
In the pure triplet case, we need to make a simplifying assumption about the nature of the pairing to analyze the state purity due to the possible $\bm{\k}$ dependence of the $\dv$ vector direction. Assuming that the direction of $\dv(\k)$ is fixed (with potentially varying magnitude), it makes a constant polar angle $\theta_{\k} = \theta$ with a chosen photon polarization axis $\zh$. In this case, the polarization state of the photons is described by the density matrix
\begin{equation}\label{eq:triplet-polarization}
	\rho^{(t)}(\theta) = \frac{1}{2}\pmat{\sin^2\theta & 0 & 0 & 0 \\ 0 & \cos^2\theta & \cos^2\theta & 0 \\ 0 & \cos^2\theta & \cos^2\theta & 0 \\ 0 & 0 & 0 & \sin^2\theta},
\end{equation}
resulting in the following angular dependence of the purity
\begin{equation}\label{eq:triplet-purity}
	\gamma^{(t)}(\theta) = \frac{1}{2}(\sin^4\theta + 2\cos^4\theta).
\end{equation}
For a general angle $\theta$, the photon polarization state is a mixture of the entangled state $\tfrac{1}{\sqrt{2}}(\ket{RL} + \ket{LR})$ with the product states $\ket{RR},\ket{LL}$. This is due to non-zero $d_{x,y}$ components describing Cooper pairs with non-zero $\zh$ projection of total spin, which degrade the state purity. When $\dv$ is parallel to the photon polarization axis ($\theta = 0,\pi$), pure, entangled photon pairs are produced. Conversely, when $\dv$ is perpendicular to the photon axis ($\theta = \tfrac{\pi}{2}$), only photons with product state polarizations can be produced. The angular dependence of the state purity is shown in Figs.~\ref{fig:purity}(a) and~\ref{fig:purity}(b). 

Finally, we arrive at the present case of mixed singlet-triplet pairing relevant to the P-N-S heterostructure. We first consider the case of a $\dv$ vector with fixed direction\cite{puetter2012identifying}, and an induced singlet component $\psi(\k)$ through Rashba SOC. To simplify the analysis we assume that the induced singlet component is $s$-wave, so that the ``fraction'' of pairing in the singlet and triplet channels, Eq.~(\ref{eq:pairing-fraction}), is independent of $\k$:
\begin{equation}
	s \approx \frac{|\Delta_0|}{\sqrt{\norm{\dv}^2 + |\Delta_0|^2}}, \qquad t^2 = 1 - s^2.
\end{equation}
Since the induced singlet component has magnitude on the order of $\lambda$, the ratio of $t$ and $s$ depends on the ratio between the Rashba SOC strength and the superconducting gap $\norm{\dv}$. The polarization state of the photons is described by the density matrix 
\begin{equation}\label{eq:mixed-polarization}
	\hspace{-2mm}\rho^{(m)}(\theta) = \frac{1}{2}\pmat{t^2\sin^2\theta & 0 & 0 & 0 \\ 0 & s^2 + t^2\cos^2\theta & s^2 + t^2\cos^2\theta & 0 \\ 0 & s^2 + t^2\cos^2\theta & s^2 + t^2\cos^2\theta & 0 \\ 0 & 0 & 0 & t^2\sin^2\theta},
\end{equation}
leading to a state purity with angular dependence
\begin{equation}\label{eq:mixed-purity}
	\gamma^{(m)}(\theta) = \frac{1}{2}[t^4\sin^4\theta + 2(s^2 + t^2\cos^2\theta)^2].
\end{equation}
This is one of the central results of this work. Note that in the limit $t \rightarrow 0$ we recover the singlet result, $\gamma = 1$, while in the limit $t \rightarrow 1$ we recover the pure triplet result. Distribution of the purity for $t = 0.77$ is shown in Figs.~\ref{fig:purity}(c) and~\ref{fig:purity}(d). The effect of the singlet component in the order parameter is to enhance the production of entangled photons. As such, the peak around $\theta = 0,\pi$ where $\gamma \simeq 1$ is broadened, while the purity around $\theta = \tfrac{\pi}{2}$ is reduced. 

\begin{figure}
	\includegraphics[width=0.85\linewidth]{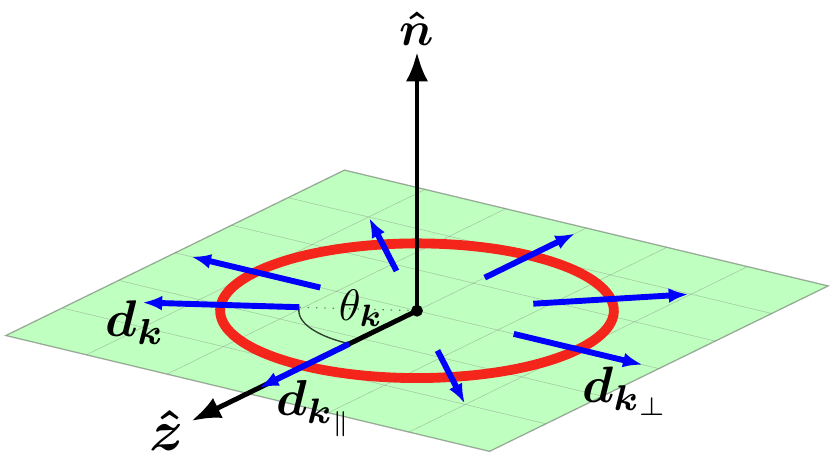}
	\caption{(colour online) Illustration of the induced $\dv$ vector from $s$-wave pairing due to Rashba SOC, shown as blue arrows on the Fermi surface in red. When the photon axis lies in the plane of the heterostructure, there are two parallel $\dv$ vectors at $\pm\k_{\parallel}$ , and two perpendicular at $\pm\k_{\perp}$. All Cooper pairs, except those with $\pm\k_{\parallel}$, make a finite angle $\theta_{\k}$ with $\zh$, and thus degrade the state purity.}
	\label{fig:mixing}
\end{figure}

We now consider the case of an $s$-wave superconductor, with induced triplet component $\dv(\k)$ through Rashba SOC. As we have seen, $\dv(\k) \propto \bm{\hat{k}}$, whose direction lies in the plane of the heterostructure, as shown in Fig.~\ref{fig:mixing}. Due to the directional dependence of $\dv(\k)$ on $\k$, the angle $\theta_{\k}$ it makes with a chosen photon axis is not constant, which means that the density matrix~(\ref{eq:mixed-density-matrix}) can only be obtained by performing the sum over $\k$. Despite this, we can still talk about the purity of the two-photon state. For simplicity, we consider $\zh$ in the plane of the heterostructure, as shown in Fig.~\ref{fig:mixing}. There are two wave vectors $\pm \k_{\parallel}$ where $\dv(\k)$ is parallel to $\zh$ ($\theta_{\k} = 0,\pi$), and those Cooper pairs can recombine to produce entangled photons. However, there are also wave vectors $\pm\k_{\perp}$ where $\dv(\k)$ is perpendicular to $\zh$ ($\theta_{\k} = \tfrac{\pi}{2}$), and those Cooper pairs can recombine to produce photons with product state polarizations. In fact, all $\dv(\k)$ except those with $\pm \k_{\parallel}$ make a finite angle $\theta_{\k}$ with $\zh$, so the Cooper pairs have finite spin projection onto the photon polarization axis. It is this finite spin projection which degrades the purity of the photon state, and we conclude that pure, entangled photons cannot be produced. More generally, if $\dv(\k)$ changes direction with $\k$, there is no choice of photon axis $\zh$ which is parallel to all $\dv(\k)$, so pure, entangled photons cannot be produced.

\section{Summary and Discussion\label{sec:discussion}}

In summary, we studied the effect of spin-triplet Cooper pairs on the state purity of photon pairs produced in a P-N-S heterostructure. A unique directional dependence of the state purity was found, which can be understood from recombination of a pure spin-triplet Cooper pair with a fixed $\dv$ vector. If the photon polarization axis $\zh$ is chosen parallel to $\dv$, then the Cooper pair has zero total spin with respect to $\zh$, as in Eq.~(\ref{eq:spin-projection}). The spin state of electrons in the Cooper pair is entangled in the $m_z = 0$ state, $\ket{\uparrow \downarrow}_z + \ket{\downarrow \uparrow}_z$, which translates into entanglement of the photon polarizations, Eq.~(\ref{eq:photon-state}), through the recombination process. If the photon polarization axis is not parallel to  $\dv$, then the triplet Cooper pair has net spin projection onto $\zh$. This means the spin state of electrons in the Cooper pair contains $m_z = \pm 1$ components, $\ket{\uparrow\uparrow}_z$, $\ket{\downarrow\downarrow}_z$, which generate product polarization states of the photons, $\ket{RR},\ket{LL}$. We find that pure, entangled photons are produced when $\zh\parallel\dv$ ($\theta = 0,\pi$, where $\theta$ is the angle between $\zh$ and $\dv$), which is diminished away from that direction. 

A singlet-triplet mixture of Cooper pairs is generically present in a P-N-S heterostructure, which is an effect due to Rashba SOC. If one starts with an $s$-wave superconductor, the induced triplet Cooper pairs are described by $\bm{d}(\k) \propto \bm{\hat{k}}$, which lies in the plane of the heterostructure. Due to the directional dependence of $\dv(\k)$ on $\k$, there is no choice of photon polarization axis $\zh$ which is parallel to all $\dv(\k)$. As such, there always exist Cooper pairs with finite spin projection onto a chosen $\zh$ ($\theta_{\k} \ne 0,\pi$), and these non-vanishing $m_z = \pm 1$ components generate product state polarizations of the photons. Therefore, pure, entangled photons cannot be produced. Since the magnitude of the induced triplet component is on the order of $\lambda$, this effect could be suppressed by selecting materials with small Rashba SOC strength relative to the singlet gap $|\Delta_0|$.

On the other hand, if one starts with a spin-triplet superconductor with fixed $\dv$, the induced singlet Cooper pairs from Rashba SOC can only enhance the production of entangled photons. This is due to the fact that a spin singlet is rotationally invariant and has zero spin projection, $m_z = 0$, on any photon polarization axis. As we have seen, the singlet component tends to broaden the peak of the purity around $\theta = 0,\pi$ where $\gamma \simeq 1$ and reduces the purity around $\theta = \tfrac{\pi}{2}$. Since the magnitude of the induced singlet component is also on the order of $\lambda$, small Rashba SOC relative to the underlying triplet gap will lead to a tight angular distribution of pure, entangled photons along $\dv$.

These considerations may aid in the design of functional devices, based on P-N-S junctions, as they offer a direction of maximum purity of entangled photon sources inherent to the heterostructure.

\begin{acknowledgments}
This work was supported by the Natural Sciences and Engineering Research Council of Canada and the Center for
Quantum Materials at the University of Toronto. 
\end{acknowledgments}\vspace{-0.5mm}

\appendix
\setcounter{secnumdepth}{0}
\section{Appendix}
Starting with the second-order time-evolved state~(\ref{eq:time-evolved-state}), the leading-order expression for the density matrix~(\ref{eq:density-matrix-definition}) involves four powers of the interaction Hamiltonian~(\ref{eq:interaction-H})\vspace{-0.5mm}
\begin{align}
\begin{split}
	\rho_{\alpha\beta\gamma\delta}(\qa,\qb) &= \int_{-\infty}^t\d t_1\int_{-\infty}^{t_1}\d t_2\int_{-\infty}^{t}\d t_3\int_{-\infty}^{t_3}\d t_4\ \times \\
	&\hspace{-10mm}\braket{\Psi_0|\V(t_2)\V(t_1)\ad{\qa,\alpha}\ad{\qb,\beta}\a{\qa,\gamma}\a{\qb,\delta}\V(t_3)\V(t_4)|\Psi_0}.
\end{split}
\end{align}
The resulting function $F(\k,\qa,\qb,\Delta_{\k})$ differs slightly from the original work of \citet{hayat2014cooper}, which can be traced back to incorrect operator ordering of the interaction Hamiltonian in the time-evolved bra above. Inserting the form of the interaction, we can use the fact that the initial state $\ket{\Psi_0} = \ket{0}\otimes\ket{\mathrm{FS}}\otimes\ket{\mathrm{BCS}}$ is a product of photon, hole, and electron sectors to compute the matrix elements separately. \par 

Starting with the photon sector, the time dependence of the operators contribute
\begin{equation}\label{eq:photon-time}
	e^{-i\w_{\q_1}t_1}e^{-i\w_{\q_2}t_2}e^{+i\w_{\q_3}t_3}e^{+i\w_{\q_3}t_3},
\end{equation}
while there are four non-zero contractions 
\begin{align}\label{eq:photon-contractions}
\begin{split}
&\braket{\a{\q_2,\s_2}\a{\q_1,\s_1}\ad{\qa,\alpha}\ad{\qb,\beta}\a{\qa,\gamma}\a{\qb,\delta}\ad{\q_3,\s_3}\ad{\q_4,\s_4}} \\
&\hspace{2mm}= \delta_{\s_1,\alpha}\delta_{\q_1,\qa}\delta_{\s_2,\beta}\delta_{\q_2,\qb}\delta_{\s_3,\gamma}\delta_{\q_3,\qa}\delta_{\s_4,\delta}\delta_{\q_4,\qb} + \\
&\hspace{6.5mm} \delta_{\s_1,\alpha}\delta_{\q_1,\qa}\delta_{\s_2,\beta}\delta_{\q_2,\qb}\delta_{\s_3,\delta}\delta_{\q_3,\qb}\delta_{\s_4,\gamma}\delta_{\q_4,\qa} + \\
&\hspace{6.5mm} \delta_{\s_1,\beta}\delta_{\q_1,\qb}\delta_{\s_2,\alpha}\delta_{\q_2,\qa}\delta_{\s_3,\gamma}\delta_{\q_3,\qa}\delta_{\s_4,\delta}\delta_{\q_4,\qb} + \\
&\hspace{6.5mm} \delta_{\s_1,\beta}\delta_{\q_1,\qb}\delta_{\s_2,\alpha}\delta_{\q_2,\qa}\delta_{\s_3,\delta}\delta_{\q_3,\qb}\delta_{\s_4,\gamma}\delta_{\q_4,\qa},
\end{split}
\end{align}
where the label $i\in\{1,2,3,4\}$ labels the momentum and angular momentum sum in $\V(t_i)$.  From the hole sector we have time dependence
\begin{equation}\label{eq:hole-time} 
e^{+i\eh_{\q_1-\k_1}t_1}e^{+i\eh_{\q_2-\k_2}t_2}e^{-i\eh_{\q_3-\k_3}t_3}e^{-i\eh_{\q_4-\k_4}t_4},
\end{equation}
and two non-zero contractions,
\begin{align}\label{eq:hole-contractions}
\begin{split}
&\braket{\hd{\q_2 - \k_2,-J_2}\hd{\q_1-\k_1,-J_1}\h{\q_3-\k_3,-J_3}\h{\q_4-\k_4,-J_4}} \\
&\hspace{2mm}= f^p_{\k_1-\q_1}f^p_{\k_2-\q_2}\delta_{J_1,J_3}\delta_{\q_1-\k_1,\q_3-\k_3}\delta_{J_2,J_4}\delta_{\q_2-\k_2,\q_4-\k_4} -\\
&\hspace{6.5mm} f^p_{\k_1-\q_1}f^p_{\k_2-\q_2}\delta_{J_1,J_4}\delta_{\q_1-\k_1,\q_4-\k_4}\delta_{J_2,J_3}\delta_{\q_2-\k_2,\q_3-\k_3},
\end{split}
\end{align}
where $f_{\k}^h = f(\eh_{\k})$ is the Fermi-Dirac distribution of holes. As mentioned earlier, the energy of the photons are on the order of the semiconductor band gap. This is typically on the order of electron volts, yielding photons with wave vector much smaller than the Fermi wave vector. Therefore, in the calculation we make the approximation that the $\k$-dependent quantities vary slowly on the scale of $\q$, and we replace $\k_i \pm \q_{A,B}$ by $\k_i$. \par 

In the electron sector we encounter expectations of the form $\braket{\cd{\k_2,j_2}\cd{\k_1,j_1}\c{\k_3,j_3}\c{\k_4,j_4}}$, where $j_i = J_i + \s_i$. The superconducting contribution to the density matrix is the contraction $\braket{\cd{\k_2,j_2}\cd{\k_1,j_1}}\braket{\c{\k_3,j_3}\c{\k_4,j_4}}$, which is non-vanishing in the superconducting state when $\k_1 = -\k_2$ and $\k_3 = -\k_4$. These terms describe a second-order process in which a Cooper pair is destroyed. Focusing on the general case of a spin singlet-triplet mixture, we use $\hat{v}_{\k}^{(m)}$ in the Bogoliubov transformation Eq.~(\ref{eq:bogoliubov-transformation}) to evaluate the matrix elements. Gathering the results into matrix form, we find
\begin{widetext}
\begin{align}\label{eq:cdcd-contractions}
\begin{split}
\left(\braket{\cd{\k_2,j_2}(t_2)\cd{\k_1,j_1}(t_1)}\right) &= e^{+i\mu_nt_1}e^{+i\mu_nt_2}\delta_{\k_1,-\k_2}u_{\k_2}v_{\k_2}\left[e^{-iE_{\k_2}t_1}e^{+iE_{\k_2}t_2}f_{\k_2}^n - e^{+iE_{\k_2}t_1}e^{-iE_{\k_2}t_2}(1 - f_{\k_2}^n)\right]\times \\
&\hspace{47mm}\pmat{-t_{\k_2}\sin\theta_{\k_2} e^{+i\vp_{\k_2}} & t_{\k_2}\cos\theta_{\k_2} + s_{\k_2} \\ t_{\k_2}\cos\theta_{\k_2} - s_{\k_2} & +t_{\k_2}\sin\theta_{\k_2} e^{-i\vp_{\k_2}}},
\end{split}
\end{align}
\begin{align}\label{eq:cc-contractions}
\begin{split}
\left(\braket{\c{\k_3,j_3}(t_3)\c{\k_4,j_4}(t_4)}\right) &= e^{-i\mu_nt_3}e^{-i\mu_nt_4}\delta_{\k_3,-\k_4}u_{\k_4}v_{\k_4}\left[e^{-iE_{\k_4}t_3}e^{+iE_{\k_4}t_4}(1 - f_{\k_4}^n) - e^{+iE_{\k_4}t_3}e^{-iE_{\k_4}t_4}f_{\k_4}^n\right]\times \\
&\hspace{47mm}\pmat{-t_{\k_4}\sin\theta_{\k_4} e^{-i\vp_{\k_4}} & t_{\k_4}\cos\theta_{\k_4} + s_{\k_4} \\ t_{\k_4}\cos\theta_{\k_4} - s_{\k_4} & +t_{\k_4}\sin\theta_{\k_4} e^{+i\vp_{\k_4}}},
\end{split}
\end{align}
\end{widetext}
where $f_{\k}^n = f(E_{\k})$ is the Fermi-Dirac distribution of Bogoliubov quasiparticles. The limits $t_{\k}\rightarrow 0$ and $s_{\k}\rightarrow 0$ yield matrix elements for the pure singlet and triplet cases, respectively. In calculating the density matrix elements, the above terms always appear in the combination $\braket{\cd{\k,j_1}\cd{-\k,j_2}}\braket{\c{-\k,j_2}\c{\k,j_1}}$, neglecting the small $\q_{A,B}$. So the angle- and mixture-dependent quantities in the above matrices appear in the combinations
\begin{align}
\begin{split}
	(s_{\k} \pm t_{\k}\cos\tk)(s_{\k} \pm t_{\k}\cos\tk)^* &= s_{\k}^2 + t_{\k}^2\cos^2\tk \\
														   &\hspace{8mm} \pm 2s_{\k}t_{\k}\cos\tk, \\
	(t_{\k}\sin\tk e^{\pm i\pk})(t_{\k}\sin\tk e^{\pm i\pk})^* &= t_{\k}^2\sin^2\tk.
\end{split}
\end{align}
The cross term $\pm 2s_{\k}t_{\k}\cos\tk$ will vanish upon summation over $\k$ because all other quantities are even under $\k \mapsto -\k$, while $\cos\tk$ flips sign due to the odd orbital part of the pairing. This fact leads to the density matrix~(\ref{eq:mixed-density-matrix}) which is a weighted sum of the pure singlet and triplet density matrices at each $\k$. 

Once the matrix elements have been computed, we need to perform the time integrals appearing in the second-order perturbation theory, which are of the form
\begin{equation}
  \int_{-\infty}^t\d t_1\int_{-\infty}^{t_1}\d t_2 e^{-i\Omega_1 t_1}e^{-i\Omega_2 t_2},
\end{equation}
multiplied by
\begin{equation}
 \int_{-\infty}^t\d t_3\int_{-\infty}^{t_3}\d t_4 e^{+i\Omega_3 t_3}e^{+i\Omega_4 t_4},
\end{equation}
where 
\begin{align}
\begin{split}
	\Omega_1 &= E_{\q_1,\k_1} + \vp E_{\k_1} \qquad \Omega_3 = E_{\q_3,\k_3} + \vp' E_{\k_3}, \\
	\Omega_2 &= E_{\q_2,\k_2} - \vp E_{\k_2} \qquad \Omega_4 = E_{\q_4,\k_4} - \vp' E_{\k_4},
\end{split}
\end{align}
$E_{\q,\k} \coloneqq \w_{\q} - \eh_{\k} - \mu_n$, and $\vp,\vp' = \pm 1$. We introduce a convergence factor $\eta$ through 
\begin{equation}
	\Omega_{1,2} \rightarrow \Omega_{1,2} + i\eta, \qquad \Omega_{3,4} \rightarrow \Omega_{3,4} - i\eta, \vspace{2mm}
\end{equation}
which serves to adiabatically turn on the interaction $\V(t)$ from $t_0 \rightarrow -\infty$. The integrals evaluate to 
\begin{equation}
	\frac{1}{(\Omega_2 + i\eta)(\Omega_4 - i\eta)}\frac{e^{4\eta t}}{[(\Omega_1 + \Omega_2)^2 + (2\eta)^2]},
\end{equation}
having used conservation of energy $\Omega_1 + \Omega_2 = \Omega_3 + \Omega_4$. The quantity computed is the pair production probability, which we must differentiate to obtain the production rate\cite{landau2013quantum}. Using the Lorentzian representation of the $\delta$ function, in the limit $\eta \rightarrow 0^+$ the production rate becomes
\begin{equation}
	\frac{2\pi\delta(E_{\q_1,\k_1} + E_{\q_2,\k_2})}{(E_{\q_2,\k_2} - \vp E_{\k_2})(E_{\q_4,\k_4} - \vp' E_{\k_4})},
\end{equation}
in which the $\delta$ function will always evaluate to $\delta(\w_{\qa} + \w_{\qb} - 2\eh_{\k} - 2\mu_n)$.
Using the matrix elements and the form of the time integrals, the function $F(\k,\qa,\qb,\Delta_{\k})$ describing the rate of pair production evaluates to
\begin{widetext}
\begin{align}\label{eq:production-rate}
\begin{split}
	F(\k,\qa,\qb,\Delta_{\k}) &= 4\pi|B_{\k,\qa}|^2|B_{\k,\qb}|^2\left|\frac{\Delta_{\k}}{2E_{\k}}\right|^2(f_{\k}^h)^2\delta(\w_{\qa} + \w_{\qb} - 2\eh_{\k} - 2\mu_n)\biggr[ \frac{f_{\k}^nf_{\k}^n}{(E_{\k} - E_{\qa,\k})^2} + \frac{(1 - f_{\k}^n)(1 - f_{\k}^n)}{(E_{\k} + E_{\qa,\k})^2} + \\
&\hspace{-15mm}   2\frac{f_{\k}^n(1 - f_{\k}^n)}{(E_{\k} - E_{\qa,\k})(E_{\k} + E_{\qa,\k})}  + \frac{f_{\k}^nf_{\k}^n}{(E_{\k} - E_{\qa,\k})(E_{\k} - E_{\qb,\k})} +  \frac{(1 - f_{\k}^n)(1 - f_{\k}^n)}{(E_{\k} + E_{\qa,\k})(E_{\k} + E_{\qb,\k})} + \\
&\frac{f_{\k}^n(1 - f_{\k}^n)}{(E_{\k} + E_{\qa,\k})(E_{\k} - E_{\qb,\k})} + \frac{f_{\k}^n(1 - f_{\k}^n)}{(E_{\k} - E_{\qa,\k})(E_{\k}+ E_{\qb,\k})} + (\qa \leftrightarrow \qb)\biggr], 
\end{split}
\end{align}
\end{widetext}
where $(\qa\leftrightarrow\qb)$ represents another seven terms with $\qa$ and $\qb$ exchanged. 

\bibliography{references}

\end{document}